\documentclass[twocolumn,showpacs,preprintnumbers,amsmath,amssymb,pre]{revtex4}

\usepackage{graphicx}
\usepackage{dcolumn}
\usepackage{bm}
\usepackage{enumerate}

\begin{document}

\title{Boundary conditions at a thin membrane that generate non--Markovian normal diffusion}

\author{Tadeusz Koszto{\l}owicz}
 \email{tadeusz.kosztolowicz@ujk.edu.pl}
 \affiliation{Institute of Physics, Jan Kochanowski University,\\
         \'Swi\c{e}tokrzyska 15, 25-406 Kielce, Poland}

\date{\today}

\begin{abstract}
We show that some boundary conditions assumed at a thin membrane may result in normal diffusion not being the stochastic Markov process. We consider boundary conditions defined in terms of the Laplace transform in which there is a linear combination of probabilities and probability fluxes defined on both membrane surfaces. The coefficients of the combination may depend on the Laplace transform parameter. Such boundary conditions are most commonly used when considering diffusion in a membrane system unless collective or non-local processes in particles diffusion occur. We find Bachelier-Smoluchowski-Chapmann-Kolmogorov (BSCK) equation in terms of the Laplace transform and we derive the criterion to check whether the boundary conditions lead to fundamental solutions of diffusion equation satisfying this equation. If the BSCK equation is not met, the Markov property is broken. When a probability flux is continuous at the membrane, the general forms of the boundary conditions for which the fundamental solutions meet the BSCK equation are derived. A measure of broken of semi-group property is also proposed. The relation of this measure to the non-Markovian property measure is discussed.
\end{abstract}

\maketitle

\section{Introduction\label{Sec1}}

A lot of processes in biology and physics are based on diffusion occurring in membrane systems in which a thin membrane is represented by a partially permeable wall \cite{hobbie,luckey,hsieh}. We mention here diffusion of substances through the skin \cite{schumm}, in the brain \cite{zhan}, and between blood and a cell \cite{kim}. 
Various kinds of diffusion are considered as stochastic processes which can be classified according to different criteria, such as non--stationarity, non--ergodicity, non--Gaussianity, ageing property, Markovian property and others \cite{mjcb,bgm,ks}. Weak ergodicity is breaking when the ensemble and time averaged mean square displacement of a particle are different. Ageing properties are defined as dependence of physical observables on the time difference between initialisation of the process and the start of the measurement. A diffusion process is Markovian when the conditional probability density of finding a particle at the point $x$ at time $t$, provided that it was at points $x'_1,\ldots,x'_{n-1},x'_n$ in the earlier moments $t'_1<\ldots<t'_{n-1}<t'_n$ satisfies the equation $P(x,t|x'_n,t'_n;x'_{n-1},t'_{n-1};\ldots;\ldots;x'_1,t'_1)\equiv P(x,t|x'_n,t'_n)$. 

The nature of diffusion is characterized by its properties and certain functions. An example is the relation $\left\langle (\Delta x (t))^2\right\rangle\sim f(t)$, where $\left\langle (\Delta x)^2\right\rangle$ means square displacement of a diffusing particle, which is often used to define normal or anomalous diffusion; when $f(t)=t^\alpha$ normal diffusion is for $\alpha=1$, superdiffusion for $\alpha>1$ and subdiffusion for $0<\alpha<1$, when $f(t)$ is a slowly varying function, such as a logarithmic function, we have slow subdiffusion (ultraslow diffusion). However, a proper combination of subdiffusion and superdiffusion processes leads to the relation characterized normal diffusion although the process is non-Markovian and non-Gaussian in nature \cite{dgn}. Thus, one needs to consider more properties mentioned above to characterize a diffusion process. One of the most important features is the Markov property. Although van Kampen mentioned that `non-Markov is the rule, Markov is the exception' \cite{vk2}, it is very often assumed that a considered process is Markovian, at least `approximately'. This is due mainly to practical reasons, since a Markov process is relatively easy to model. Namely, this process is fully determined by both the conditional probability $P(x,t|x',t')$ and the probability describing initial state $P(x_0,t_0)$ only. If the process is Markovian, then the conditional probability fulfils the Bachelier-Smoluchowski-Chapmann-Kolmogorov (BSCK) equation \cite{gardiner,risken,vk1}
\begin{equation}\label{eq1}
P(x,t|x_0,t_0)=\int_{-\infty}^\infty dx' P(x,t|x',t')P(x',t'|x_0,t_0).
\end{equation}
Thus, if Eq. (\ref{eq1}) is not met, there is a non-Markovian process. In other words, the semi-group property is broken \cite{fel1}. However, if Eq. (\ref{eq1}) is met, it is not obvious if the process is Markovian \cite{af1,af2,wf}. 

An example of a non-Markovian diffusion is subdiffusion. Subdiffusion occurs in media in which particle jumps are strongly hindered due to a complex structure of the medium. The example are subdiffusion in gels \cite{kdm,bm}, biological cells \cite{bgm,bf}, membranes \cite{mjc}, and in media having a fractal structure \cite{iom}. Within the Continuous Time Random Walk model waiting time for the particle to jump is anomalously long for subdiffusion; the probability density distribution of this time $\psi$ has a heavy tail, $\psi(t)\sim 1/t^{\alpha+1}$, $t\rightarrow\infty$, $0<\alpha<1$, which leads to infinite mean value of this time \cite{ks,mk}. Subdiffusion in a homogeneous system can be described by a differential equation with the Riemann-Liouville time derivative of a fractional order $\partial P(x,t)/\partial t=D_\alpha\partial^{1-\alpha}/\partial t^{1-\alpha}(\partial^2 P(x,t)/\partial x^2)$, $0<\alpha<1$, where $D_\alpha$ is a subdiffusion coefficient. The Riemann--Liouville fractional derivative is defined for $\beta>0$ as 
\begin{equation}\label{eq2}
\frac{d^\beta f(t)}{dt^\beta}=\frac{1}{\Gamma(n-\beta)}\frac{d^n}{dt^n}\int_0^t dt'(t-t')^{n-\beta-1}f(t'),
\end{equation}
where $n=[\beta]+1$, $[\beta]$ is the integral part of $\beta$. Power-law distribution $\psi$ with a heavy tail leads to ageing of the system as well as to WEB. The presence of a fractional derivative in the subdiffusion equation shows that subdiffusion is not a Markovian process unlike normal diffusion. 

For normal diffusion in a homogeneous system the average waiting time for the particle to jump is finite, $\left\langle \tau\right\rangle=\int_0^\infty \psi(\tau)d\tau<\infty$, providing the process to be Markovian, ergodic, and free of ageing features. However, some factors may change the properties. Weak ergodicity breaking of normal diffusion can be observed in heterogeneous medium \cite{ccm}. Anomalous diffusion can emerge from ergodicity breaking \cite{mgp}. It has been shown that far from equilibrium transport of a periodically driven inertial particle moving in a periodic potential within a classical Markovian dynamics with Brownian motion provides ergodicity breaking without the need to introduce heavy-tailed distributions \cite{slh}. There are processes in systems with normal diffusion in which an `obstacle' has been located. Examples of this are diffusion in a system with subdiffusive membrane \cite{mjc,cc} and diffusion of an antibiotic in a system with a bacterial biofilm \cite{km,kmwa}. It has been shown \cite{kwl} that the Riemann--Liouville fractional time derivative of the $1/2$ order is involved in boundary conditions at a partially permeable wall for normal diffusion. The question arises how the presence of such an `obstacles' changes the properties of normal diffusion.

An important issue is to find methods that allow one to check property of diffusion processes. A distinction between normal diffusion and subdiffusion can be made by means of the single particle tracking method \cite{slr}, fluorescence recovery after photobleaching method \cite{amb}, the method based on temporal evolution of near--membrane layers \cite{kdm}, and others. Identifying the Markov property from experimental results is a more difficult task. In practice, this is only possible in some processes, such as ion current flowing through membrane channels \cite{ful1}. Various measures of deviation from the non-Markov property have been proposed. The considerations mainly concern quantum systems \cite{ckr}, see also \cite{lg} and the references cited therein. The proposed methods are not equivalent, they are often based on the interpretation of the Markov process. The example are the measure based on the failure of the semi--group property and on a quantum information flow \cite{lu}. 

In this paper, we consider diffusion of a particle in a system with a thin membrane. We derive criteria for checking whether fundamental solutions meet the BSCK equation. We also propose a measure of how far fundamental solutions are from satisfying the BSCK equation. We also discuss whether this measure can be taken as a measure of broken Markov property. We find the general form of fundamental solutions to normal diffusion equation for these conditions, and then we derive criteria to check whether the solutions also meet the BSCK equation. Failure of the last equation shows that the process is non--Markovian. 

Modelling the diffusion process in a membrane system is convenient to perform the considerations in terms of the Laplace transform $\mathcal{L}[f(t)]\equiv\int_0^\infty {\rm e}^{-st}f(t)dt\equiv\hat{f}(s)$. The main results are presented in terms of the Laplace transform.

The organization of this paper is as follows. In Sec. \ref{Sec2} we derive the BSCK equation in terms of the Laplace transform. In Sec. \ref{Sec3} we discuss various forms of boundary conditions at a thin membrane and the general forms of the fundamental solutions to diffusion equation obtained for these boundary conditions. We consider Laplace transforms of boundary conditions as a linear combination of probabilities describing diffusion of a single particle and probability fluxes defined on both membrane surfaces, with coefficient depending on the Laplace transform parameter. Such boundary conditions are often used when considering the diffusion in a membrane system. We consider local boundary conditions for the diffusion equation, the permeability membrane properties do not change over time. In Sec. \ref{Sec4} we derive criteria that allow us to check whether fundamental solutions meet the BSCK equation. Final remarks and conclusions are presented in Sec. \ref{Sec5}.

\section{Bachelier-Smoluchowski-Chapmann-Kolmogorov equation in terms of the Laplace transform\label{Sec2}}

Normal diffusion in a system with constant diffusion coefficient $D$ is the Wiener stationary process. The conditional probability density of finding a diffusing particle at the point $x$ at time $t$ under condition that at the initial moment $t_0$ the particle was at the position $x_0$ depends on the time difference \cite{gardiner,risken}
\begin{equation}\label{eq3}
P(x,t|x_0,t_0)\equiv P(x,t-t_0|x_0). 
\end{equation}
In further considerations we assume $t_0=0$. The process is described by the normal diffusion equation
\begin{equation}\label{eq4}
\frac{\partial P(x,t|x_0)}{\partial t}=D\frac{\partial^2 P(x,t|x_0)}{\partial x^2},
\end{equation}
the initial condition is $P(x,0|x_0)=\delta(x-x_0)$, where $\delta$ is the Dirac delta function. The solution to Eq. (\ref{eq4}) for this initial condition is called the fundamental solution.

The Laplace transform of Eq. (\ref{eq4}) reads
\begin{equation}\label{eq5}
s\hat{P}(x,s|x_0)-P(x,0|x_0)=D\frac{\partial^2 \hat{P}(x,s|x_0)}{\partial x^2}.
\end{equation}
We express Eq. (\ref{eq1}) in terms of the Laplace transform. Using Eq. (\ref{eq3}), integrating both side of Eq. (\ref{eq1}) with respect to $t'$ in the time interval $(0,t)$ and putting $t_0=0$ we obtain
\begin{eqnarray}\label{eq6}
tP(x,t|x_0)\\
=\int_{-\infty}^\infty dx'\int_0^t dt'P(x,t-t'|x')P(x',t'|x_0),\nonumber
\end{eqnarray}
Due to the relations $\mathcal{L}[tf(t)]=-d\hat{f}(s)/ds$ and $\mathcal{L}\left[\int_0^t f(t-t')g(t')dt'\right]=\hat{f}(s)\hat{g}(s)$, we get
\begin{equation}\label{eq7}
-\frac{d\hat{P}(x,s|x_0)}{ds}=\int_{-\infty}^\infty dx'\hat{P}(x,s|x')\hat{P}(x',s|x_0).
\end{equation}
Eq. (\ref{eq7}) is the BSCK equation in terms of the Laplace transform.

\section{Boundary conditions at a thin membrane\label{Sec3}}

Boundary conditions (BCs) at a thin membrane are associated with a certain process of transporting the particle through the membrane. Such a process can be quite complex and cause a disturbance of Markov properties for diffusion of molecules located especially near the membrane. The boundary conditions with respect to normal diffusion have been often derived by means of a phenomenological model or just assumed. Membrane boundary conditions which are not equivalent to one another have been presented in \cite{zhang,goychuk,tk,tk1,ab}. 

Nonlocal boundary conditions, such as Wentzell--Neumann one \cite{ism}, are not considered here. We assume that permeability membrane properties do not change over time. Diffusing particles are not accumulated inside a membrane and do not clog a membrane. Assuming that particles diffuse independently of one another, the same boundary conditions can be used for probability densities describing single particle diffusion as well as for concentrations of the particles. 

\begin{figure}[h]
        \includegraphics[scale=0.6]{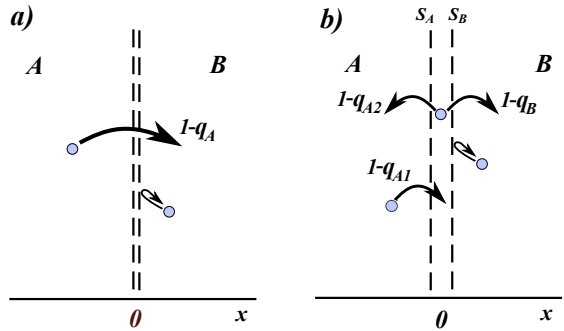}
				\caption{Diffusion in a system with a partially absorbing membrane. A particle that has passed from region $A$ to $B$ cannot return to the region $A$. In Fig. a) the particle cannot stop inside the membrane, in Fig. b) the particle can do it. $S_A$ and $S_B$ denote membrane surfaces, the parameters $1-q_i$ next to the arrows are the probability of passing the particle through the surfaces.\label{Fig1}}
\end{figure}

\begin{figure}[h]
        \includegraphics[scale=0.6]{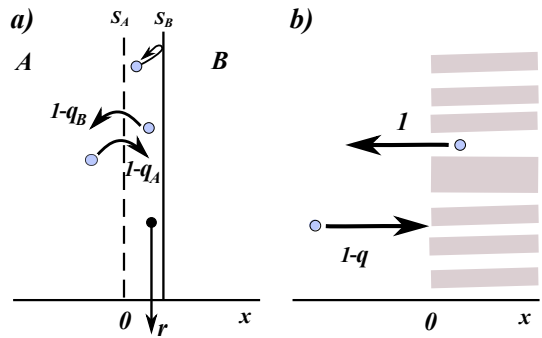}
				\caption{In Fig. a) diffusion in a system with a partially absorbing wall is shown. The particle can jump into the membrane and be absorbed, $r$ is the absorption parameter. The particle cannot penetrate into region $B$. In Fig. b) a fully one-sided permeable surface is at $x=0$. A particle that tries to jump over the surface by jumping from the right side of the system to the left one can do it without any obstacle.\label{Fig2}}
\end{figure}

\begin{figure}[h]
        \includegraphics[scale=0.6]{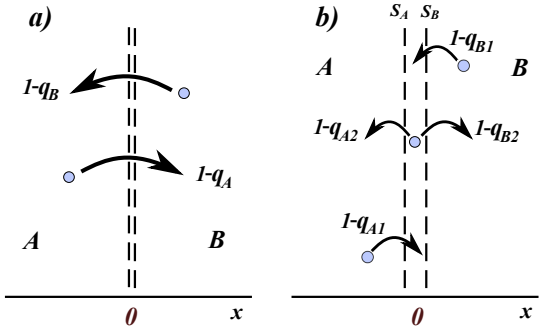}
				\caption{Situation similar to the one presented in Fig. \ref{Fig1} but for a partially permeable membrane for particles moving in both directions.\label{Fig3}}
\end{figure}

When deriving the boundary conditions at the membrane, we have postulated the following rule \cite{kd}: If the diffusion equation is derived from a certain theoretical model, the boundary conditions at the thin membrane can also be derived from this model with additional assumptions taking into account the selective properties of the membrane. Examples illustrating boundary conditions are shown in Figs. \ref{Fig1}--\ref{Fig3}. The symbols $1-q_i$ next to the arrows mean the probabilities of passing of a diffusing particle through the membrane surface, $q_i$ is the probability that particle is stopped at the membrane surface.  We assume that the membrane is thin enough that it can be treated as a partially permeable wall or absorbing wall located at $x=0$. The particle may, with some probability, jump into the membrane and back again, but its diffusion inside the membrane is not possible. We mention that if the membrane were sufficiently thick, the model could be extended. Namely, we could consider diffusion in a three-layer system. The middle part would represent a membrane inside which diffusion of particles takes place. 

Particle random walk models on a discrete lattice are effective at deriving boundary conditions at the border between media. Some models assume that there is a point at the boundary between media or inside the membrane at which the molecule must be stopped temporarily \cite{goychuk}, see Figs. \ref{Fig1}b), \ref{Fig2}a), and \ref{Fig3}b). In another model, it is assumed that the molecule can jump across the border between the media without having to stop at the border \cite{tk,tk1}, see Figs. \ref{Fig1}a), \ref{Fig2}b), and \ref{Fig3}a). Both models can lead to different boundary conditions.

In the following we mark the function $P$ by the indexes $i$ and $j$ which indicate the location of the points $x$ and $x_0$, respectively. Assuming that the thin membrane is placed at $x=0$, the indexes $i$ and $j$ denote the signs of $x$ and $x_0$, respectively. In the time domain the diffusive fluxes are defined as
\begin{equation}\label{eq8}
J_{ij}(x,t|x_0)=-D\frac{\partial P_{ij}(x,t|x_0)}{\partial x},
\end{equation}
$i,j\in\{-,+\}$. In terms of the Laplace transform Eq. (\ref{eq8}) reads
\begin{equation}\label{eq9}
\hat{J}_{ij}(x,s|x_0)=-D\frac{\partial \hat{P}_{ij}(x,s|x_0)}{\partial x}.
\end{equation}

\subsection{Diffusion in a half-space\label{Sec3a}}

We consider diffusion in a region $A=(-\infty,0)$ bounded by a thin membrane located at $x=0$. Two boundary conditions are needed to solve the diffusion equation. One condition reads 
\begin{equation}\label{eq10}
\hat{P}_{--}(-\infty,s|x_0)=0, 
\end{equation}
and the other is assumed at the membrane.

\subsubsection{Partially absorbing wall}

We suppose that the boundary condition takes the form
\begin{equation}\label{eq11}
\hat{J}_{--}(0^-,s|x_0)=\hat{\Phi}(s)\hat{P}_{--}(0^-,s|x_0).
\end{equation}
In the time domain this boundary condition reads
\begin{equation}\label{eq12}
J_{--}(0^-,t|x_0)=\int_0^t dt'\Phi(t-t')P_{--}(0^-,t'|x_0).
\end{equation}
The fundamental solution to Eq. (\ref{eq5}) for boundary conditions Eqs. (\ref{eq10}) and (\ref{eq11}) is
\begin{eqnarray}\label{eq13}
\hat{P}_{--}(x,s|x_0)=\frac{1}{2\sqrt{Ds}}\Bigg[{\rm e}^{-|x-x_0|\sqrt{\frac{s}{D}}}\\
+\frac{\sqrt{Ds}-\hat{\Phi}(s)}{\sqrt{Ds}+\hat{\Phi}(s)}{\rm e}^{(x+x_0)\sqrt{\frac{s}{D}}}\Bigg].\nonumber
\end{eqnarray}

In Figs. \ref{Fig1} and \ref{Fig2} a) the models of partially absorbing wall are shown. In Fig. \ref{Fig1} a particle that is placed initially in the region $A=(-\infty,0)$ after moving to the region $B=(0,\infty)$ cannot return to $A$. 
In Fig. \ref{Fig1}a) a particle can jump through the membrane with $1-q$ probability without being able to stop inside the membrane. In this case we get \cite{tk}
\begin{equation}\label{eq14}
\hat{J}_{--}(0^-,s|x_0)=\lambda\hat{P}_{--}(0^-,s|x_0),
\end{equation}
where $\lambda$ is a constant coefficient controlled by the probability $1-q$. In the time domain we have $\Phi(t)=\lambda\delta(t)$ which leads to the Robin boundary condition
\begin{equation}\label{eq15}
J_{--}(0^-,t|x_0)=\lambda P_{--}(0^-,t|x_0).
\end{equation}

The processes presented in Figs. \ref{Fig1}b) and \ref{Fig2}a) lead to the function $\hat{\Phi}(s)$ that may depend explicitly on the parameter $s$. In Fig. \ref{Fig1}b) a particle can stop temporarily in the membrane and can return to the region $A$ with probability $1-q_{A2}$ or can jump to the region $B$ with probability $1-q_B$; a particle that is in $B$ cannot escape from this region. The dwell time of the molecule inside the membrane depends on the probabilities mentioned above as well as the distribution $\psi_M$ of the waiting time for a jump of the particle located inside the membrane. In Fig. \ref{Fig2}a) a particle can be eliminated from further diffusion, with the probability controlled by a parameter $r$, due to absorption or irreversible chemical reactions. We suppose that $\hat{\Phi}(s)$ may be complicated function of $s$, especially for multistage absorption process near an impenetrable surface \cite{chou}.

\subsubsection{Fully reflecting or fully absorbing wall}

The boundary condition at a fully reflecting wall is $J_{--}(0^-,t|x_0)=0$ which provides $\hat{\Phi}(s)\equiv 0$, and BC for fully absorbing wall is $P_{--}(0^-,t|x_0)=0$ which leads to $\hat{\Phi}(s)\equiv\infty$. The fundamental solution reads
\begin{equation}\label{eq16}
\hat{P}_{--}(x,s|x_0)=\frac{1}{2\sqrt{Ds}}\Bigg[{\rm e}^{-|x-x_0|\sqrt{\frac{s}{D}}}
\pm {\rm e}^{(x+x_0)\sqrt{\frac{s}{D}}}\Bigg],
\end{equation}
where sign $+$ before the last term is for fully reflecting wall and $-$ for fully absorbing wall.

\subsection{Diffusion in an unbounded system\label{Sec3b}}

We assume that diffusion is considered in the regions $A$ and $B$ simultaneously. To solve Eq. (\ref{eq5}) in both regions separated by the membrane one needs four boundary conditions. Two of the boundary conditions read 
\begin{equation}\label{eq17}
\hat{P}_{-\pm}(-\infty,s|x_0)=\hat{P}_{+\pm}(\infty,s|x_0)=0,
\end{equation}
and two others are fixed at the membrane. 

The boundary conditions at an asymmetrical membrane should be different depending on which part of the system the diffusing particle is located initially. In order to justify this statement let us consider diffusion of two particles $U1$ and $U2$ located symmetrically with respect to the membrane at the initial moment; their probability distributions are denoted as $P_{U1}(x,t;x_0)$ and $P_{U2}(x,t;-x_0)$, respectively. The probabilities of finding the particle $U1$ in the region $x<0$ and the particle $U2$ in the region $x>0$ at time $t>0$ cannot be equal when the membrane is asymmetrical, $\int_{-\infty}^0\hat{P}_{U1}(x,s|x_0)dx\neq\int_0^\infty\hat{P}_{U2}(x,s|-x_0)dx$. This condition is fulfilled only if at least one of the boundary conditions at the membrane is different for the particles $U1$ and $U2$. Thus, boundary conditions are defined separately for the cases of $x_0<0$ and $x_0>0$. 

We assume that the boundary conditions at the membrane in terms of the Laplace transform are as follows
\begin{equation}\label{eq18}
\hat{P}_{+-}(0^+,s|x_0)=\hat{\Phi}_1(s)\hat{P}_{--}(0^-,s|x_0),
\end{equation}
\begin{equation}\label{eq19}
\hat{J}_{+-}(0^+,s|x_0)=\hat{\Xi}_1(s)\hat{J}_{--}(0^-,s|x_0),
\end{equation}
for $x_0<0$ and
\begin{equation}\label{eq20}
\hat{P}_{-+}(0^-,s|x_0)=\hat{\Phi}_2(s)\hat{P}_{++}(0^+,s|x_0),
\end{equation}
\begin{equation}\label{eq21}
\hat{J}_{-+}(0^-,s|x_0)=\hat{\Xi}_2(s)\hat{J}_{++}(0^+,s|x_0),
\end{equation}
for $x_0>0$. We assume that $\hat{\Phi}_{i}(s)\;,\;\hat{\Xi}_{i}(s)\geq 0$, $i=1,2$. In the time domain the boundary conditions read
\begin{equation}\label{eq18a}
P_{+-}(0^+,t|x_0)=\int_0^t dt'\Phi_1(t-t')P_{--}(0^-,t'|x_0),
\end{equation}
\begin{equation}\label{eq19a}
J_{+-}(0^+,t|x_0)=\int_0^t dt'\Xi_1(t-t')J_{--}(0^-,t'|x_0),
\end{equation}
for $x_0<0$ and
\begin{equation}\label{eq20a}
P_{-+}(0^-,t|x_0)=\int_0^t dt'\Phi_2(t-t')P_{++}(0^+,t'|x_0),
\end{equation}
\begin{equation}\label{eq21a}
J_{-+}(0^-,t|x_0)=\int_0^t dt'\Xi_2(t-t')J_{++}(0^+,t'|x_0),
\end{equation}
for $x_0>0$.

The fundamental solutions to Eq. (\ref{eq5}) for the boundary conditions Eqs. (\ref{eq17})--(\ref{eq21}) are
\begin{eqnarray}\label{eq22}
\hat{P}_{--}(x,s|x_0)=\frac{1}{2\sqrt{Ds}}\;{\rm e}^{-|x-x_0|\sqrt{\frac{s}{D}}}\\
-\left(\frac{\hat{\Phi}_1(s)-\hat{\Xi}_1(s)}{\hat{\Phi}_1(s)+\hat{\Xi}_1(s)}\right)\frac{1}{2\sqrt{Ds}}\;{\rm e}^{(x+x_0)\sqrt{\frac{s}{D}}}\nonumber,
\end{eqnarray}
\begin{eqnarray}\label{eq23}
\hat{P}_{+-}(x,s|x_0)\\
=\left(\frac{\hat{\Phi}_1(s)\hat{\Xi}_1(s)}{\hat{\Phi}_1(s)+\hat{\Xi}_1(s)}\right)
\frac{1}{\sqrt{Ds}}\;{\rm e}^{-(x-x_0)\sqrt{\frac{s}{D}}}\nonumber,
\end{eqnarray}
\begin{eqnarray}\label{eq24}
\hat{P}_{-+}(x,s|x_0)\\
=\left(\frac{\hat{\Phi}_2(s)\hat{\Xi}_2(s)}{\hat{\Phi}_2(s)+\hat{\Xi}_2(s)}\right)
\frac{1}{\sqrt{Ds}}\;{\rm e}^{-(x_0-x)\sqrt{\frac{s}{D}}}\nonumber,
\end{eqnarray}
\begin{eqnarray}\label{eq25}
\hat{P}_{++}(x,s|x_0)=\frac{1}{2\sqrt{Ds}}\;{\rm e}^{-|x-x_0|\sqrt{\frac{s}{D}}}\\
-\left(\frac{\hat{\Phi}_2(s)-\hat{\Xi}_2(s)}{\hat{\Phi}_2(s)+\hat{\Xi}_2(s)}\right)\frac{1}{2\sqrt{Ds}}\;{\rm e}^{-(x+x_0)\sqrt{\frac{s}{D}}}\nonumber.
\end{eqnarray}

Below are considered examples of boundary conditions. For simplicity, we assume that $x_0<0$.

\subsubsection{Partially permeable wall}

The random walk model in a system with a thin membrane, applying for the system presented in Fig. \ref{Fig3}a) (a particle cannot be stopped inside the membrane), for $0<q_A,q_B<1$, provides \cite{tk}
\begin{equation}\label{eq26}
\hat{\Phi}_1(s)=\frac{1}{a+b\sqrt{s}},
\end{equation}
$a,b>0$ are parameters controlled by the probabilities $q_A$ and $q_B$. The second boundary condition is that the flux is continuous at the membrane 
\begin{equation}\label{eq27}
J_{--}(0^-,t|x_0)=J_{+-}(0^+,t|x_0)\equiv J(x,t|x_0), 
\end{equation}
which provides $\hat{\Xi}_1(s)\equiv 1$. 

From Eqs. (\ref{eq22}), (\ref{eq23}), and (\ref{eq26}) we obtain the following Laplace transforms of fundamental solutions
\begin{eqnarray}\label{eq32}
\hat{P}_{--}(x,s|x_0)=\frac{1}{2\sqrt{Ds}}\;{\rm e}^{-|x-x_0|\sqrt{\frac{s}{D}}}\\
-\left(\frac{1-a-b\sqrt{s}}{1+a+b\sqrt{s}}\right)\frac{1}{2\sqrt{Ds}}\;{\rm e}^{(x+x_0)\sqrt{\frac{s}{D}}}\nonumber,
\end{eqnarray}
\begin{eqnarray}\label{eq33}
\hat{P}_{+-}(x,s|x_0)\\
=\left(\frac{1}{1+a+b\sqrt{s}}\right)
\frac{1}{\sqrt{Ds}}\;{\rm e}^{-(x-x_0)\sqrt{\frac{s}{D}}}\nonumber.
\end{eqnarray}

Using the formula
\begin{equation}\label{eq28}
\mathcal{L}\left[\frac{d^\beta f(t)}{dt^\beta}\right]=s^\beta\hat{f}(s),
\end{equation}
$0<\beta<1$, from Eqs. (\ref{eq18}) and (\ref{eq26}) we get
\begin{equation}\label{eq29}
P_{--}(0^-,t|x_0)=a P_{+-}(0^+,t|x_0)+b\frac{\partial^{1/2}P_{+-}(0^+,t|x_0)}{\partial t^{1/2}}.
\end{equation}
Thus, the boundary condition involves the Riemann--Liouville fractional time derivative of the $1/2$ order. Calculating the inverse Laplace transform of Eq. (\ref{eq29}) , the BC may be presented in the following form
\begin{equation}\label{eq30}
P_{+-}(0^+,t|x_0)=\int_0^t F(t-t')P_{--}(0^-,t'|x_0)dt',
\end{equation}
where
\begin{equation}\label{eq31}
F(t)=\frac{1}{b}\left[\frac{1}{\sqrt{Dt}}-\frac{a}{b}\;{\rm e}^{\frac{a^2t}{b^2}}{\rm erfc}\left(\frac{a\sqrt{t}}{b}\right)\right],
\end{equation}
${\rm erfc}(u)\equiv (2/\sqrt{\pi})\int_u^\infty {\rm e}^{-\tau^2}d\tau$ is the complementary error function. 

Let us consider the following BC at the membrane
\begin{equation}\label{eq34}
J(0,t|x_0)=\lambda_1 P_{--}(0^-,t|x_0)-\lambda_2 P_{+-}(0^+,t|x_0),
\end{equation}
$\lambda_1,\lambda_2>0$. The fundamental solutions to the diffusion equation Eq. (\ref{eq5}) for Laplace transforms of BCs Eqs. (\ref{eq27}) and (\ref{eq34}) are
\begin{eqnarray}\label{eq35}
\hat{P}_{--}(x,s|x_0)=\frac{1}{2\sqrt{Ds}}\;{\rm e}^{-|x-x_0|\sqrt{\frac{s}{D}}}\\
-\left(\frac{\lambda_1-\lambda_2-\sqrt{Ds}}{\lambda_1+\lambda_2+\sqrt{Ds}}\right)\frac{1}{2\sqrt{Ds}}\;{\rm e}^{(x+x_0)\sqrt{\frac{s}{D}}}\nonumber,
\end{eqnarray}
\begin{eqnarray}\label{eq36}
\hat{P}_{+-}(x,s|x_0)\\
=\left(\frac{\lambda_1}{\lambda_1+\lambda_2+\sqrt{Ds}}\right)
\frac{1}{\sqrt{Ds}}\;{\rm e}^{-(x-x_0)\sqrt{\frac{s}{D}}}\nonumber.
\end{eqnarray}
Eqs. (\ref{eq32}) and (\ref{eq33}) are identical to Eqs. (\ref{eq35}) and (\ref{eq36}), respectively, if $a=\lambda_2/\lambda_1$ and $b=\sqrt{D}/\lambda_1$. Because the boundary conditions uniquely determine the solutions to the diffusion equation, we conclude that the boundary conditions (\ref{eq29}), (\ref{eq30}), and (\ref{eq34}) are equivalent to each other. Thus, we have shown that the boundary condition Eq. (\ref{eq29}) can be written in two other equivalent forms that do not contain a fractional time derivative. This example also shows that the boundary conditions expressed in the form of a linear combination of probabilities and fluxes can be represented in the form of Eqs. (\ref{eq18}) and (\ref{eq19}), at least when the flux is continuous at the membrane.

\subsubsection{One-sided fully permeable wall}

For the situation presented in Fig. \ref{Fig2}b) the random walk model provides the flux continuity $\hat{\Xi}_1(s)\equiv 1$ and \cite{tk}
\begin{equation}\label{eq38}
\hat{\Phi}_1(s)=\frac{1-q_A}{s}.
\end{equation}

In the examples presented above, it was assumed that the particle cannot stop inside the membrane. However, if we assume that such a retention of the particle is possible, the continuity of the flux could be broken. In such cases, the functions $\hat{\Phi}_i(s)$, $\hat{\Xi}_i(s)$, $i=1,2$,  may take more complicated forms.

\section{When fundamental solutions meet BSCK equation\label{Sec4}}

The conditions checking if fundamental solutions meet the BSCK equation are derived separately for diffusion in half-space and in an unbounded system. 

\subsection{Diffusion in a half space\label{Sec4a}}

We consider diffusion of a particle in the region $A=(-\infty,0)$. The BSCK equation in this region reads
\begin{equation}\label{eq39}
-\frac{d\hat{P}_{--}(x,s|x_0)}{ds}=\int_{-\infty}^0 dx'\hat{P}_{--}(x,s|x')\hat{P}_{--}(x',s|x_0).
\end{equation}
$x,x_0<0$. 
The function $R_{--}$ is defined by its Laplace transform as follows
\begin{eqnarray}\label{eq40}
\hat{R}_{--}(x,s|x_0)=-\frac{d\hat{P}_{--}(x,s|x_0)}{ds}\\
-\int_{-\infty}^0 dx'\hat{P}_{--}(x,s|x')\hat{P}_{--}(x',s|x_0).\nonumber
\end{eqnarray}

\subsubsection{Partially absorbing wall}

From Eqs. (\ref{eq13}) and (\ref{eq40}) we get
\begin{eqnarray}\label{eq41}
\hat{R}_{--}(x,s|x_0)=\frac{\hat{\Phi}'(s)}{(\sqrt{Ds}+\hat{\Phi}(s))^2}{\rm e}^{(x+x_0)\sqrt{\frac{s}{D}}},
\end{eqnarray}
$x,x_0<0$, where $\hat{\Phi}'(s)=d\hat{\Phi}(s)/ds$. The function Eq. (\ref{eq13}) fulfils BSCK equation if 
\begin{equation}\label{eq42}
\hat{R}_{--}(x,s|x_0)=0,
\end{equation}
which provides $\hat{\Phi}(s)=\kappa=const$. In the time domain we get
\begin{equation}\label{eq43}
\Phi(t)\equiv\kappa\delta(t).
\end{equation}

Thus, if the integral operator kernel in a boundary condition is time-dependent, i.e. the boundary condition is not given as $J(0^-,t|x_0)=\kappa P(0^-,t|x_0)$ with a constant $\kappa$, the BSCK equation is not met and the diffusion process is non-Markovian. 

\subsubsection{Fully reflecting or fully absorbing wall}

For diffusion in systems with fully reflecting wall or fully absorbing wall fundamental solution Eq. (\ref{eq16}) fulfils Eq. (\ref{eq42}).

\subsection{Diffusion in an unbounded system\label{Sec4b}}

Using the notation of fundamental solutions defined in this paper, Eq. (\ref{eq7}) takes the following form
\begin{eqnarray}\label{eq44}
-\frac{d\hat{P}_{ij}(x,s|x_0)}{ds}\\
=\int_{-\infty}^{0} dx'\hat{P}_{i-}(x,s|x')\hat{P}_{-j}(x',s|x_0)\nonumber\\
+\int_{0}^\infty dx'\hat{P}_{i+}(x,s|x')\hat{P}_{+j}(x',s|x_0)\nonumber,
\end{eqnarray}
$i,j\in\{-,+\}$.
We define the function $R_{ij}(x,t|x_0)$ by means of its Laplace transform
\begin{eqnarray}\label{eq45}
\hat{R}_{ij}(x,s|x_0)=-\frac{d\hat{P}_{ij}(x,s|x_0)}{ds}\\
-\int_{-\infty}^{0} dx'\hat{P}_{i-}(x,s|x')\hat{P}_{-j}(x',s|x_0)\nonumber\\
-\int_{0}^\infty dx'\hat{P}_{i+}(x,s|x')\hat{P}_{+j}(x',s;|_0)\nonumber.
\end{eqnarray}
From Eqs. (\ref{eq22})--(\ref{eq25}) and (\ref{eq45}) we get
\begin{eqnarray}\label{eq46}
\hat{R}_{--}(x,s|x_0)=\frac{1}{2\sqrt{Ds^3}}\;{\rm e}^{(x+x_0)\sqrt{\frac{s}{D}}}\times\nonumber\\
\Bigg[\frac{\hat{\Phi}_1(s)\hat{\Xi}_1(s)}{\hat{\Phi}_1(s)+\hat{\Xi}_1(s)}\left(\frac{1}{\hat{\Phi}_1(s)+\hat{\Xi}_1(s)}
-\frac{\hat{\Phi}_2(s)\hat{\Xi}_2(s)}{\hat{\Phi}_2(s)+\hat{\Xi}_2(s)}\right)\nonumber\\
+2s\frac{\hat{\Phi}'_1(s)\hat{\Xi}_1(s)-\hat{\Phi}_1(s)\hat{\Xi}'_1(s)}{(\hat{\Phi}_1(s)+\hat{\Xi}_1(s))^2}\Bigg],
\end{eqnarray}
\begin{eqnarray}\label{eq47}
\hat{R}_{+-}(x,s|x_0)=\frac{1}{2\sqrt{Ds^3}}\;{\rm e}^{-(x-x_0)\sqrt{\frac{s}{D}}}\times\nonumber\\
\Bigg[\frac{\hat{\Phi}_1(s)\hat{\Xi}_1(s)}{\hat{\Phi}_1(s)+\hat{\Xi}_1(s)}\Bigg(\frac{\hat{\Phi}_1(s)}{\hat{\Phi}_1(s)+\hat{\Xi}_1(s)}
-\frac{\hat{\Xi}_2(s)}{\hat{\Phi}_2(s)+\hat{\Xi}_2(s)}\Bigg)\nonumber\\
-2s\frac{\hat{\Phi}'_1(s)\hat{\Xi}^2_1(s)+\hat{\Phi}^2_1(s)\hat{\Xi}'_1(s)}{(\hat{\Phi}_1(s)+\hat{\Xi}_1(s))^2}\Bigg],
\end{eqnarray}
\begin{eqnarray}\label{eq48}
\hat{R}_{-+}(x,s|x_0)=\frac{1}{2\sqrt{Ds^3}}\;{\rm e}^{-(x_0-x)\sqrt{\frac{s}{D}}}\times\nonumber\\
\Bigg[\frac{\hat{\Phi}_2(s)\hat{\Xi}_2(s)}{\hat{\Phi}_2(s)+\hat{\Xi}_2(s)}\Bigg(\frac{\hat{\Phi}_2(s)}{\hat{\Phi}_2(s)+\hat{\Xi}_2(s)}
-\frac{\hat{\Xi}_1(s)}{\hat{\Phi}_1(s)+\hat{\Xi}_1(s)}\Bigg)\nonumber\\
-2s\frac{\hat{\Phi}'_2(s)\hat{\Xi}^2_2(s)+\hat{\Phi}^2_2(s)\hat{\Xi}'_2(s)}{(\hat{\Phi}_2(s)+\hat{\Xi}_2(s))^2}\Bigg],
\end{eqnarray}
\begin{eqnarray}\label{eq49}
\hat{R}_{++}(x,s|x_0)=\frac{1}{2\sqrt{Ds^3}}\;{\rm e}^{-(x+x_0)\sqrt{\frac{s}{D}}}\times\nonumber\\
\Bigg[\frac{\hat{\Phi}_2(s)\hat{\Xi}_2(s)}{\hat{\Phi}_2(s)+\hat{\Xi}_2(s)}\left(\frac{1}{\hat{\Phi}_2(s)+\hat{\Xi}_2(s)}
-\frac{\hat{\Phi}_1(s)\hat{\Xi}_1(s)}{\hat{\Phi}_1(s)+\hat{\Xi}_1(s)}\right)\nonumber\\
+2s\frac{\hat{\Phi}'_2(s)\hat{\Xi}_2(s)-\hat{\Phi}_2(s)\hat{\Xi}'_2(s)}{(\hat{\Phi}_2(s)+\hat{\Xi}_2(s))^2}\Bigg],
\end{eqnarray}
where $\hat{\Phi}'_i(s)=d\hat{\Phi}_i(s)/ds$ and $\hat{\Xi}'_i(s)=d\hat{\Xi}_i(s)/ds$.

The functions $\hat{\Phi}_{1}$, $\hat{\Xi}_{1}$, $\hat{\Phi}_{2}$, and $\hat{\Xi}_{2}$ provide the fundamental solutions which fulfil the BSCK equation Eq. (\ref{eq18}) if
\begin{equation}\label{eq50}
\hat{R}_{ij}(x,s|x_0)= 0,
\end{equation}
for all $i$ and $j$.
Combining the equations $\hat{R}_{--}(x,s|x_0)=0$ and $\hat{R}_{+-}(x,s|x_0)=0$ we get
\begin{equation}\label{eq51}
\frac{\hat{\Phi}_1(s)\hat{\Xi}_2(s)\left(\hat{\Phi}_1(s)\hat{\Phi}_2(s)-1\right)}{\hat{\Phi}_2(s)+\hat{\Xi}_2(s)}=2s\hat{\Phi}'_1(s),
\end{equation}
\begin{eqnarray}\label{eq52}
\frac{\hat{\Phi}_2(s)\hat{\Xi}_1(s)\left(\hat{\Phi}_1(s)+\hat{\Xi}_1(s)\right)\left(1-\hat{\Xi}_1(s)\hat{\Xi}_2(s)\right)}{\left(\hat{\Phi}_2(s)+\hat{\Xi}_2(s)\right)\left(1+\hat{\Xi}_1(s)\right)}\\
=2s\hat{\Xi}'_1(s),\nonumber
\end{eqnarray}
and from the equations $\hat{R}_{-+}(x,s|x_0)=0$ and $\hat{R}_{++}(x,s|x_0)=0$ we obtain
\begin{equation}\label{eq53}
\frac{\hat{\Phi}_2(s)\hat{\Xi}_1(s)\left(\hat{\Phi}_1(s)\hat{\Phi}_2(s)-1\right)}{\hat{\Phi}_1(s)+\hat{\Xi}_1(s)}=2s\hat{\Phi}'_2(s),
\end{equation}
\begin{eqnarray}\label{eq54}
\frac{\hat{\Phi}_1(s)\hat{\Xi}_2(s)\left(\hat{\Phi}_2(s)+\hat{\Xi}_2(s)\right)\left(1-\hat{\Xi}_1(s)\hat{\Xi}_2(s)\right)}{\left(\hat{\Phi}_1(s)+\hat{\Xi}_1(s)\right)\left(1+\hat{\Xi}_2(s)\right)}\\
=2s\hat{\Xi}'_2(s).\nonumber
\end{eqnarray}
Solutions to Eqs. (\ref{eq51})--(\ref{eq54}) can be found for some special cases only. These equations should be treated as the criterion whether the boundary conditions at the thin membrane Eqs. (\ref{eq18})--(\ref{eq21}) lead to the fundamental solutions which fulfil the BSCK equation. Below we consider three specific cases of boundary conditions at the membrane.

\subsubsection{Continuous flux at the membrane}

We assume that the flux is continuous at the membrane 
\begin{equation}\label{eq55}
\hat{J}_{-i}(0^-,s:x_0)=\hat{J}_{+i}(0^+,s:x_0),
\end{equation}
$\hat{\Xi}_1(s)=\hat{\Xi}_2(s)=1$, $i\in\{-,+\}$. Then, Eqs. (\ref{eq51}) and (\ref{eq53}) read
\begin{equation}\label{eq56}
\frac{\hat{\Phi}_1(s)\left(\hat{\Phi}_1(s)\hat{\Phi}_2(s)-1\right)}{\hat{\Phi}_2(s)+1}=2s\hat{\Phi}'_1(s),
\end{equation}
\begin{equation}\label{eq57}
\frac{\hat{\Phi}_2(s)\left(\hat{\Phi}_1(s)\hat{\Phi}_2(s)-1\right)}{\hat{\Phi}_1(s)+1}=2s\hat{\Phi}'_2(s).
\end{equation}
The solutions to Eqs. (\ref{eq56}) and (\ref{eq57}) are
\begin{equation}\label{eq58}
\hat{\Phi}_1(s)=\frac{1}{\frac{1}{\alpha}+\eta\sqrt{s}},
\end{equation}
\begin{equation}\label{eq59}
\hat{\Phi}_2(s)=\frac{1}{\alpha+\alpha\eta\sqrt{s}},
\end{equation}
where $\alpha$ and $\eta$ are constants, $\alpha>0$.
For $\eta\neq 0$, the inverse Laplace transform of Eqs. (\ref{eq18}) and (\ref{eq20}) with the kernels given by Eqs. (\ref{eq58}) and (\ref{eq59}), respectively,  provides the boundary conditions of the forms expressed by Eqs. (\ref{eq29}), (\ref{eq30}), and (\ref{eq34}). For $\eta=0$ we get $\Phi_1(t)=\alpha\delta(t)$ and $\Phi_2(t)=\delta(t)/\alpha$, then the ratio of probabilities defined at both membrane surfaces is constant.

\subsubsection{One--sided fully permeable membrane}

We consider a thin membrane that is fully permeable to particles diffusing from the region $x>0$ to the region $x<0$ and partially permeable to particles moving in the opposite direction, similar situation is presented in Fig. \ref{Fig2}b). Then, we suppose that $\hat{\Xi}_2(s)\equiv\hat{\Phi}_2(s)\equiv 1$. From Eqs. (\ref{eq51})--(\ref{eq54}) we get $\hat{\Phi}_1(s)=0$ or $\hat{\Phi}_1(s)=1$ and $\hat{\Xi}_1(s)=0$ or $\hat{\Xi}_1(s)=1$. This result means that the BSCK equation is fulfilled if the membrane is fully permeable, fully reflecting, or fully absorbing for particles diffusing from the left--hand part to right--hand part of the system. If the membrane is one-sided partially permeable we have $\hat{\Phi}_1(s)\neq 0$, $\hat{\Phi}_1(s)\neq 1$ or/and $\hat{\Xi}_2(s)\neq 0$, $\hat{\Xi}_2(s)\neq 1$. Then, Eqs. (\ref{eq51})--(\ref{eq54}) are not met and the process is non--Markovian.
 
\subsubsection{Partially absorbing membrane}

When a particle can be absorbed with a certain probability at the membrane, then $\hat{\Xi}_1(s)=\beta_1$ and/or $\hat{\Xi}_2(s)=\beta_2$, where $\beta_1$ and $\beta_2$ are constant absorption probabilities, $0<\beta_{1},\beta_{2}<1$. Assuming additionally that the membrane is not fully absorbing, $\hat{\Phi}_{1}\neq 0$ and $\hat{\Phi}_{2}\neq 0$, we find that Eqs. (\ref{eq52}) and (\ref{eq54}) are not met in this case. 

\section{Final remarks\label{Sec5}}

We have considered the normal diffusion described by Eq. (\ref{eq5}) with membrane boundary conditions Eqs. (\ref{eq18})--(\ref{eq21}).
We have shown that the fundamental solutions to the diffusion equation fulfil the BSCK equation only if the Laplace transforms of the functions $\Xi_{1}$, $\Xi_{2}$, $\Phi_{1}$ and $\Phi_{2}$, that determine the boundary conditions at a thin membrane, meet Eqs. (\ref{eq51})--(\ref{eq54}).

We have shown that boundary condition at partially absorbing wall Eq. (\ref{eq11}) leads to non-Markov process when $\hat{\Phi}$ explicitly depends on the parameter $s$. Non-Markov diffusion is also at one--sided fully permeable wall and when boundary condition, which supplements flux continuity at the membrane, is
\begin{equation}\label{eq62}
P_{--}(0^-,t|x_0)=a P_{+-}(0^+,t|x_0)+b\frac{\partial^{\alpha}P_{+-}(0^+,t|x_0)}{\partial t^\alpha}
\end{equation}
with $\alpha\neq 1/2$. 

In \cite{kwl} there has been shown the procedure of experimental derivation of boundary conditions at a thin membrane directly from experimentally obtained concentration profiles of the diffusing substance. The idea of this method is as follows. We choose functions that characterize the diffusion process in the membrane system and that can be easily determined experimentally. The example of the function is the temporal evolution of the amount of substance that diffused through the membrane from the region $A$ to $B$ (at the initial moment the substance was completely in the region $A$). The theoretically calculated functions depend on $\hat{\Xi}_1(s)$ and $\hat{\Phi}_1(s)$. Assuming that the flux is continuous at the thin membrane, $\hat{\Xi}_1(s)=1$, and comparing theoretical and experimental results, we determine the functions  $\hat{\Phi}_1(s)$. The functions have been considered in terms of the Laplace transform. Laplace transforms of experimentally determined functions have been calculated by means of the Gauss-- Laguerre quadrature and the spline interpolation method. For the case of ethanol diffusion in water, the boundary conditions at the artificial nephrophan hemodialyzer thin membrane made of cellulose acetate in time domain is given by Eq. (\ref{eq62}) with $\alpha=1/2$.

We define a measure of how far the diffusion process is from the semi--group property as
\begin{equation}\label{eq60}
R(x,t|x_0)=P(x,t|x_0)-\int_{-\infty}^\infty dx' P(x,t-t'|x')P(x',t'|x_0).
\end{equation}
From Eqs. (\ref{eq46})--(\ref{eq49}) and (\ref{eq60}) we get 
\begin{equation}\label{eq61}
R(x,t|x_0)=\mathcal{L}^{-1}\left[\sum_{i,j\in\{-,+\}}|\hat{R}_{ij}(x,s|x_0)|\right]. 
\end{equation}
For diffusion in a half-space the measure is $R(x,t|x_0)=\mathcal{L}^{-1}\left[|\hat{R}_{--}(x,s|x_0)|\right]$. If $R(x,t|x_0)\neq 0$, the diffusion process is definitely non--Markovian. 

The condition $R(x,t|x_0)\equiv 0$ does not determine whether the process is Markovian or not. If diffusion of a particle is considered in the region far from the membrane, the effect of the membrane on diffusion is negligibly small and the process is Markovian. Then, there is $|x|,|x_0|\gg 0$, so $R$ is close to zero due to the fact that the functions $\hat{R}_{ij}$ are controlled by a factor ${\rm e}^{-\sqrt{\frac{s}{D}}(|x|+|x_0|)}$. In this case the condition $R(x,t|x_0)= 0$ corresponds to the Markovian property. 

According to van Kampen's statement Markov processes are an exception. Diffusion model depends on the assumptions made. Considering different time scales in a particle random walk model, diffusion can be described by a parabolic diffusion equation, which leads to the Markov property or by a hyperbolic diffusion equation \cite{cattaneo}, which describes non-Markovian diffusion. Hyperbolic diffusion equation reads
\begin{equation}\label{eq63}
\tau \frac{\partial^2 P(x,t|x_0)}{\partial t^2}+\frac{\partial P(x,t|x_0)}{\partial t}=D\frac{\partial^2 P(x,t|x_0)}{\partial x^2},
\end{equation}
where the parameter $\tau$ is defined by means of the flux equation $J(x,t+\tau|x_0)=-D\partial P(x,t|x_0)/\partial x$; combining the flux equation with the continuity equation one gets Eq. (\ref{eq63}) in the limit of small $\tau$. Eq. (\ref{eq63}) can be derived from a persistent random walk model \cite{pottier,weiss}. In this model, the direction of a particle jump is preferred in the next step due to the inertia of the particle. The parameter $\tau$ controls this effect. If $\tau\neq 0$, the process is non-Markovian. However, if the time interval between observations of particle Brownian motion is long enough, the subsequent jumps are independent of each other. Then $\tau=0$ and the process is described by the parabolic diffusion equation as the Markov process. Usually, the easier-to-solve parabolic equation is preferred to use. However, the hyperbolic diffusion equation gives qualitatively different results than the parabolic equation even for a small parameter $\tau$ when we consider diffusion impedance \cite{kl2009} or a process in which diffusing particles can chemically react with other molecules \cite{tk2014}.
In many other cases, the solutions to the parabolic and hyperbolic diffusion equations are so close to each other for small $\tau$ that it cannot be determined from experimental studies which of them better describes the process. An example is diffusion in a homogeneous system and in a membrane system \cite{tkhyp}. In such cases, one can use a parabolic equation to describe diffusion, treating it as a Markovian approximation of the process.

An experimental checking of whether or not a process is Markovian can be carried out by various methods not equivalent to each other, often based on the interpretation of the Markov process. The measure of deviation from Markov property is, in fact, most often based on the measure of breaking the BSCK equation or equations equivalent to it such as normalized correlation functions equation \cite{af2}. As discussed in \cite{af2}, the experimental verification whether the BSCK equation is met requires a very large number of measurements of the four-dimensional matrix elements $(x,t|x',t')$. However, a large number of measurements means that the error of the determined value can be relatively large. Statistical tests can check the null hypothesis $R=0$ at some confidence level. Similar remark concerns a measure of deviation from the Markov property. We do not expect that empirical data provide the answer whether or not a process is Markovian with absolute certainty.

In our opinion, taking into account the above remarks the question should be: can a process under study be modelled as a Markov process? The answer is not obvious and this question may be treated as open. However, we put the following hypothesis: {\it if $R=0$, then diffusion in a membrane system can be modelled as a Markovian process, if there are no specific strong arguments for its non--Markovianity}. A specific strong argument may be, for example, the incompatibility of empirical data with theoretical results obtained from a model based on the assumption that the process is Markovian. We note that the occurrence of a fractional time derivative of $1/2$ order in the boundary condition at the membrane gives no reason to interpret that the diffusion process is non--Markovian. The reason is that the boundary condition can be replaced by an equivalent condition Eq. (\ref{eq34}) without a fractional derivative.

 
\section*{Acknowledgements}

The author would like to thank Prof. Andrzej Fuli\'nski for discussion and critical comments on some issues considered in this paper.

\end{document}